\documentclass[showpacs,twocolumn,prb,superscriptaddress,notitlepage,nofootinbib]{revtex4-2}
\usepackage[dvips]{graphicx}
\usepackage{amsmath,amssymb,amsthm,mathrsfs,amsfonts,dsfont}
\usepackage{subfigure, epsfig}
\usepackage{braket}
\usepackage{bm}
\usepackage{enumerate}
\usepackage{color}
\usepackage{graphicx}% Include figure files
\usepackage{dcolumn}% Align table columns on decimal point
\usepackage{bm}% bold math
\newcommand{\yms}{YbMnSb$_2$}
\usepackage[plainpages=false,colorlinks=true,linkcolor=blue,citecolor=blue,anchorcolor=blue,urlcolor=blue ]{hyperref}
\begin{document}
	\title{Magnetic and electronic structure of the topological semimetal YbMnSb$_2$}% Force line breaks with \\
\author{Jian-Rui Soh}
\affiliation{Institute of Physics, Ecole Polytechnique Fédérale de Lausanne (EPFL), CH-1015 Lausanne, Switzerland}
\author{Siobhan M. Tobin}
\affiliation{Department of Physics, University of Oxford, Clarendon Laboratory, Oxford OX1 3PU, United Kingdom}
\author{Hao Su}
\affiliation{School of Physical Science and Technology, ShanghaiTech University, Shanghai 201210, China}
\author{Ivica Zivkovic}
\affiliation{Institute of Physics, Ecole Polytechnique Fédérale de Lausanne (EPFL), CH-1015 Lausanne, Switzerland}
\author{Bachir Ouladdiaf}
\affiliation{Institut Laue-Langevin, 6 rue Jules Horowitz, BP 156, 38042 Grenoble Cedex 9, France}
\author{Anne Stunault}
\affiliation{Institut Laue-Langevin, 6 rue Jules Horowitz, BP 156, 38042 Grenoble Cedex 9, France}
\author{J. Alberto Rodr\'iguez-Velamaz\'an}
\affiliation{Institut Laue-Langevin, 6 rue Jules Horowitz, BP 156, 38042 Grenoble Cedex 9, France}
\author{Ketty Beauvois}
\affiliation{Institut Laue-Langevin, 6 rue Jules Horowitz, BP 156, 38042 Grenoble Cedex 9, France}
\author{Yanfeng Guo}
\affiliation{School of Physical Science and Technology, ShanghaiTech University, Shanghai 201210, China}
\author{Andrew T. Boothroyd}
\affiliation{Department of Physics, University of Oxford, Clarendon Laboratory, Oxford OX1 3PU, United Kingdom}
%\author{Author list}
\date{\today}% It is always \today, today,
\begin{abstract}
The antiferromagnetic (AFM) semimetal \yms\, has recently been identified as a candidate topological material, driven by time-reversal symmetry breaking. Depending on the ordered arrangement of Mn spins below the N\'{e}el temperature, $T_\mathrm{N}$ = 345\,K, the electronic bands near the Fermi energy can ether have a Dirac node, a Weyl node or a nodal line. We have investigated the ground state magnetic structure of \yms\, using unpolarized and polarized single crystal neutron diffraction.  We find that the Mn moments lie along the $c$ axis of the $P4/nmm$ space group and are arranged in a C-type AFM structure, which implies the existence of gapped Dirac nodes near the Fermi level.  The results highlight how different magnetic structures can critically affect the topological nature of fermions in semimetals.
%Our results clarify the topological nature of the charge carriers in \yms.
\begin{description}
		\item[PACS numbers]
	\end{description}
\end{abstract}
%\pacs{Valid PACS appear here}
\maketitle
\section{Introduction}

Topological semimetals are materials whose electronic bands have a linear dispersion in the vicinity of the Fermi energy. Examples include crystals where the valence and conduction bands meet at discrete points (as in Dirac or Weyl semimetals) or along a one-dimensional curve in \textbf{k}-space (in the case of nodal line semi-metals) ~\cite{Shuo2018,Fang_2016,Nitesh}. These systems can host electrons that mimic the behavior of massless fermions and are robust against perturbations due to the protection afforded by the topology of the electronic band structure~\cite{Armitage,Nagaosa2020}.

The exploration of magnetic materials with topologically non-trivial electronic band structure has become a central topic in quantum materials physics. A particular interest is to identify materials in which the topology of the electronic states can be controlled by their magnetic order~\cite{Xu2020,Zhang2021,Gui2019,WangLL2019,Soh2019c}. Recently, the antiferromagnetic (AFM) metal \yms\, has been proposed as one such material, with the property that the topology of the bands near the Fermi energy ($E_\mathrm{F}$) depends on the specific spin configuration on the Mn magnetic sublattice~\cite{Wang2018,Kealhofer2018,Qiu2019,Klemenz2019,Klemenz2020,Pan2021}. 

\begin{figure}[b!]
	\includegraphics[width=0.48\textwidth]{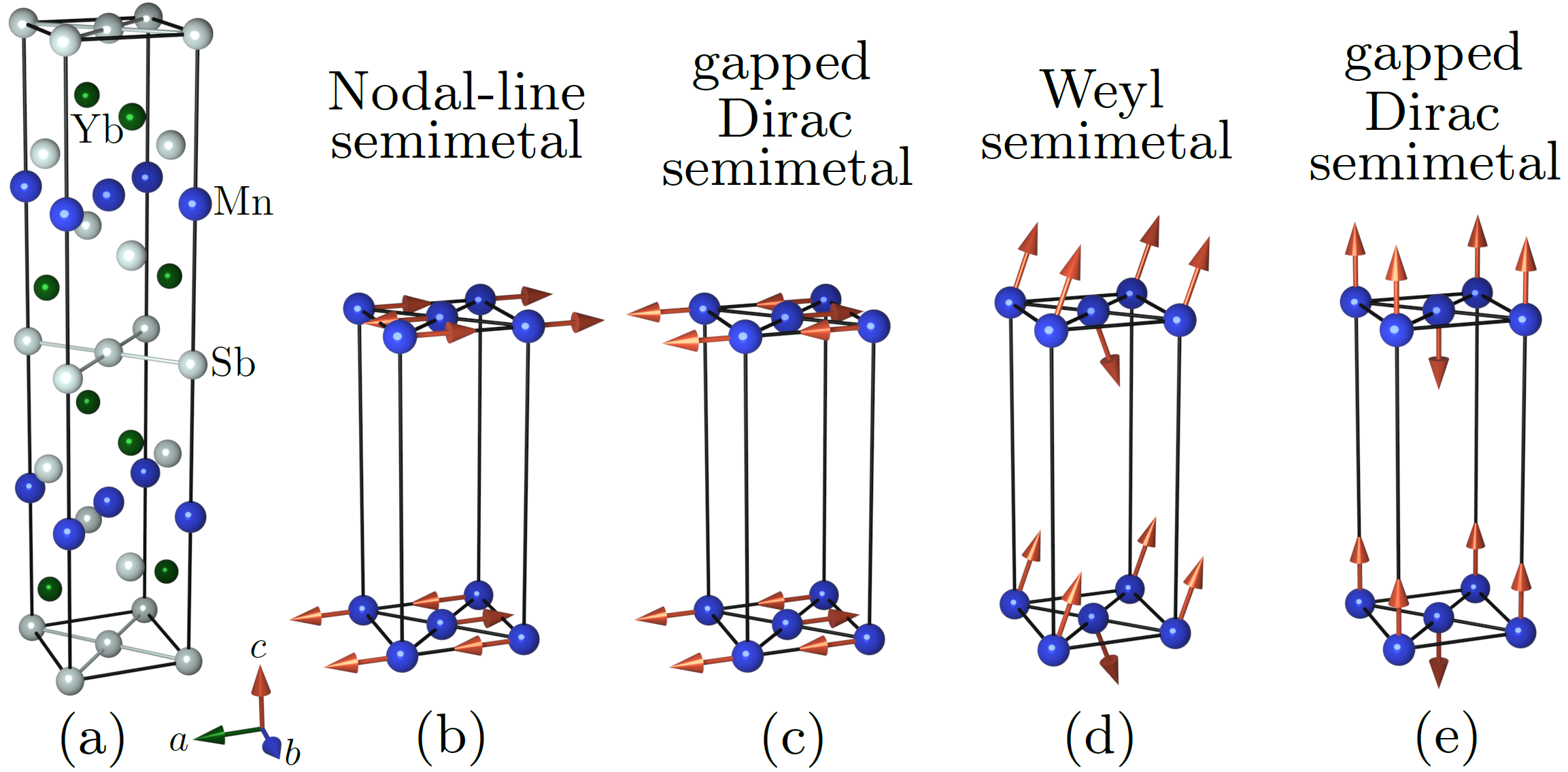}
	\caption{\label{fig:AI0_YbMnSb2_Figure_1.png} (a) Two unit cells of \yms\, as described by the space group $P4/nmm$ with cell parameters $a = b = 4.31(2)$\,\AA, $c =  10.85(1)$\,\AA\ at $T = 400$\,K. The square Sb layers (at $z =0$, $c/2$, $c$) are predicted to host different types of topological fermions, depending on the magnetic structure of the Mn magnetic sublattice. (b)--(e) show different Mn magnetic structures and the corresponding predicted topology of the electronic bands: (b) G-type AFM with in-plane Mn moments stabilizes a line node in the electronic bands; (c) C-type AFM with in-plane moments gives rise to Dirac nodes with a small gap; (d) a canted C-type AFM order gives rise to topologically protected Weyl nodes; (e) C-type AFM with moments along $c$ stabilizes a gapped Dirac state.}
\end{figure}

Previous studies have shown that the crystal structure of \yms\, can be described by the tetragonal space group $P4/nmm$ (no.~129)~\cite{Wang2018,Kealhofer2018,Qiu2019}. The unit cell, shown in Fig.~\ref{fig:AI0_YbMnSb2_Figure_1.png}(a), includes a square sublattice of Sb atoms which is predicted to host the topological fermions. This layer is in turn sandwiched between layers of Mn, facilitating the coupling between magnetism and electronic band topology. Depending on whether \yms\, displays G-type, C-type (with in-plane Mn moments), or canted C-type AFM order [see Figs.~\ref{fig:AI0_YbMnSb2_Figure_1.png}(b)--(d)], the electronic bands are predicted to give rise to a nodal-line dispersion~\cite{Qiu2019}, a gapped Dirac crossing~\cite{Kealhofer2018}, or Weyl nodes~\cite{Wang2018}, respectively. %Yet, only Weyl nodes benefit from the topological protection~\cite{Armitage,Nagaosa2020}. 
It is essential, therefore, to determine the magnetic order of the Mn sublattice in order to ascertain the topological nature of the fermions in \yms.

In this work we determined the spin configuration of the Mn magnetic  sublattice with high precision using a combination of polarized and unpolarized single crystal neutron diffraction. We find that below the N\'{e}el temperature the Mn spins point along the crystal $c$ axis and are arranged in a C-type AFM structure [Fig.~\ref{fig:AI0_YbMnSb2_Figure_1.png}(e)]. The measurements place an upper limit of 0.01\,$\mu_{\rm B}$ on any in-plane FM component (either above or below $T_{\rm N}$). The particular C-type AFM structure found here is predicted to create gapped Dirac points in the calculated band structure of \yms.
%{\color{magenta}The results will aid in the identification of the topological nature of the charge carriers in \yms. --- re-word}

\section{Methods}
Single crystalline samples of \yms\, were grown by the flux method, as described in Refs.~\cite{Wang2018,Qiu2019}, giving rise to shiny rectangular platelets with typical dimensions of $\sim\,4\times4\times0.5$ mm$^3$. The structure and quality of the single crystals were checked on a six-circle X-ray diffractometer (Oxford Diffraction) and Laue diffractometer (Multiwire Laboratories).

Magnetization measurements were performed on a superconducting quantum interference device (SQUID) magnetometer (MPMS-3, Quantum Design) with magnetic field applied parallel and perpendicular to the crystal $a$ axis. Measurements were performed in the temperature range $T$ = 2 to 400\,K and in fields of up to 3\,T. Electrical transport measurements were performed on a Physical Properties Measurement System (PPMS, Quantum Design) with the resistivity option, in the temperature range 2 to 400\,K in zero field.

\begin{figure*}[t!]
	\includegraphics[width=0.8\textwidth]{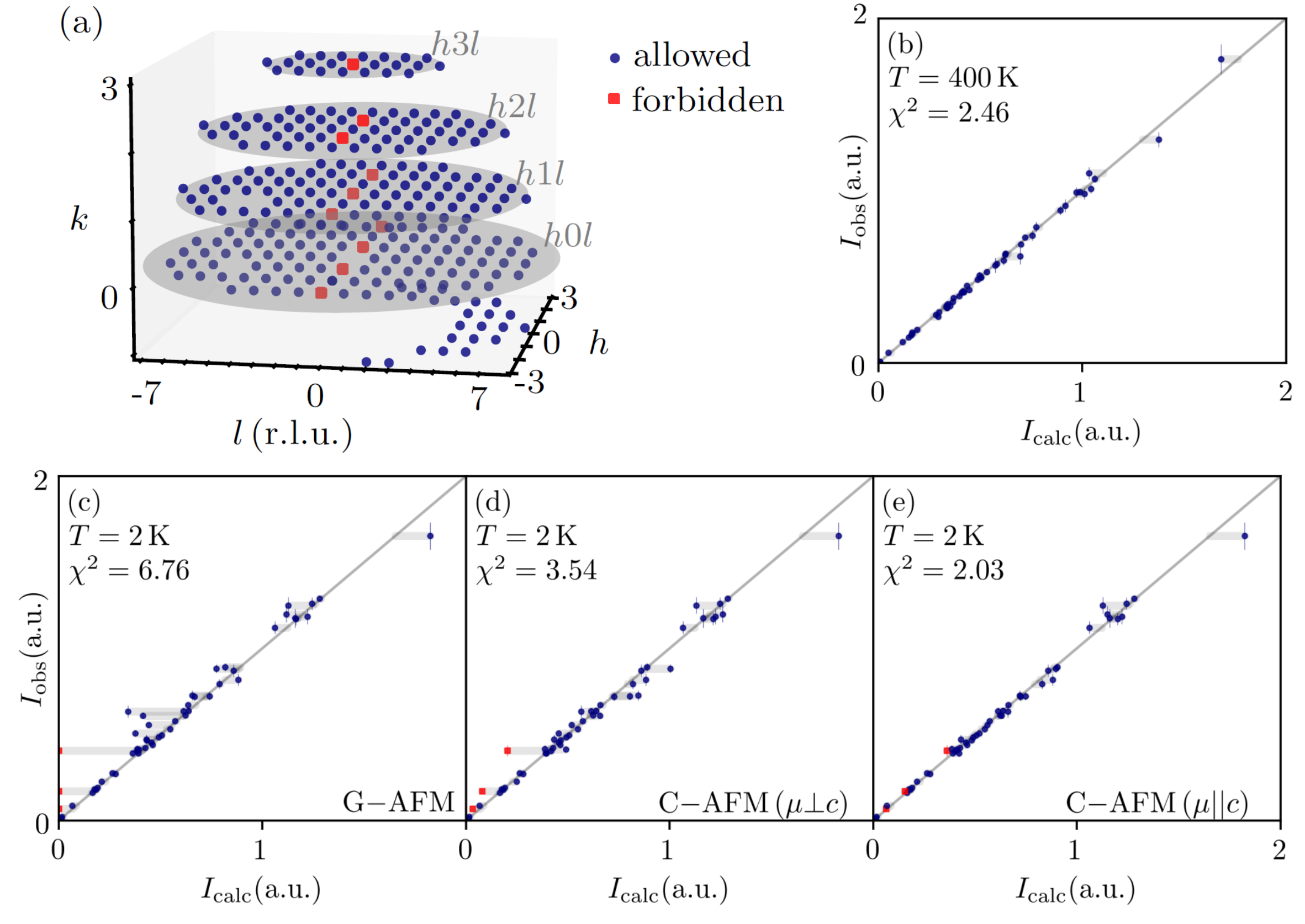}
	\caption{\label{fig:AI0_YbMnSb2_Figure_4.pdf}(a) Reciprocal space representation of the $hkl$ reflections that were measured on the D10 instrument. The blue and red points correspond  to reflections which are allowed and forbidden, respectively,  by the $P4/nmm$ space group. (b) Comparison of measured ($I_\mathrm{obs}$) and calculated ($I_\mathrm{calc}$) integrated intensities from the structural refinement in the paramagnetic phase of \yms\, at $T = 400$\,K.  (c), (d) and (e)  compare $I_\mathrm{obs}$ and $I_\mathrm{calc}$ for the $T=2$\,K data refined against structural models with G-type AFM, C-AFM (moments in-plane) and C-AFM (moments along $c$), respectively.}
\end{figure*}

Single crystal neutron diffraction with unpolarized neutrons of wavelength $\lambda = 2.36$\,{\AA} was performed on the four-circle diffractometer D10 at the Institut Laue–Langevin (ILL). The scattered neutrons were measured with a $94\times94$ mm$^2$ area detector. To determine the magnetic and crystal structure of \yms, a total of 382 and 386 $hkl$ reflections were collected below and above the magnetic ordering temperatures, at $T=2$\,K and 400\,K, respectively [Fig.~\ref{fig:AI0_YbMnSb2_Figure_4.pdf}(a)]. To ascertain if there is a canting of the Mn moments away from the crystal $c$ axis [Fig.~\ref{fig:AI0_YbMnSb2_Figure_5.png}(a),(b)], the temperature dependence several reflections were measured in the temperature range from 2 to 420\,K.
\begin{figure}[t!]
	\includegraphics[width=0.35\textwidth]{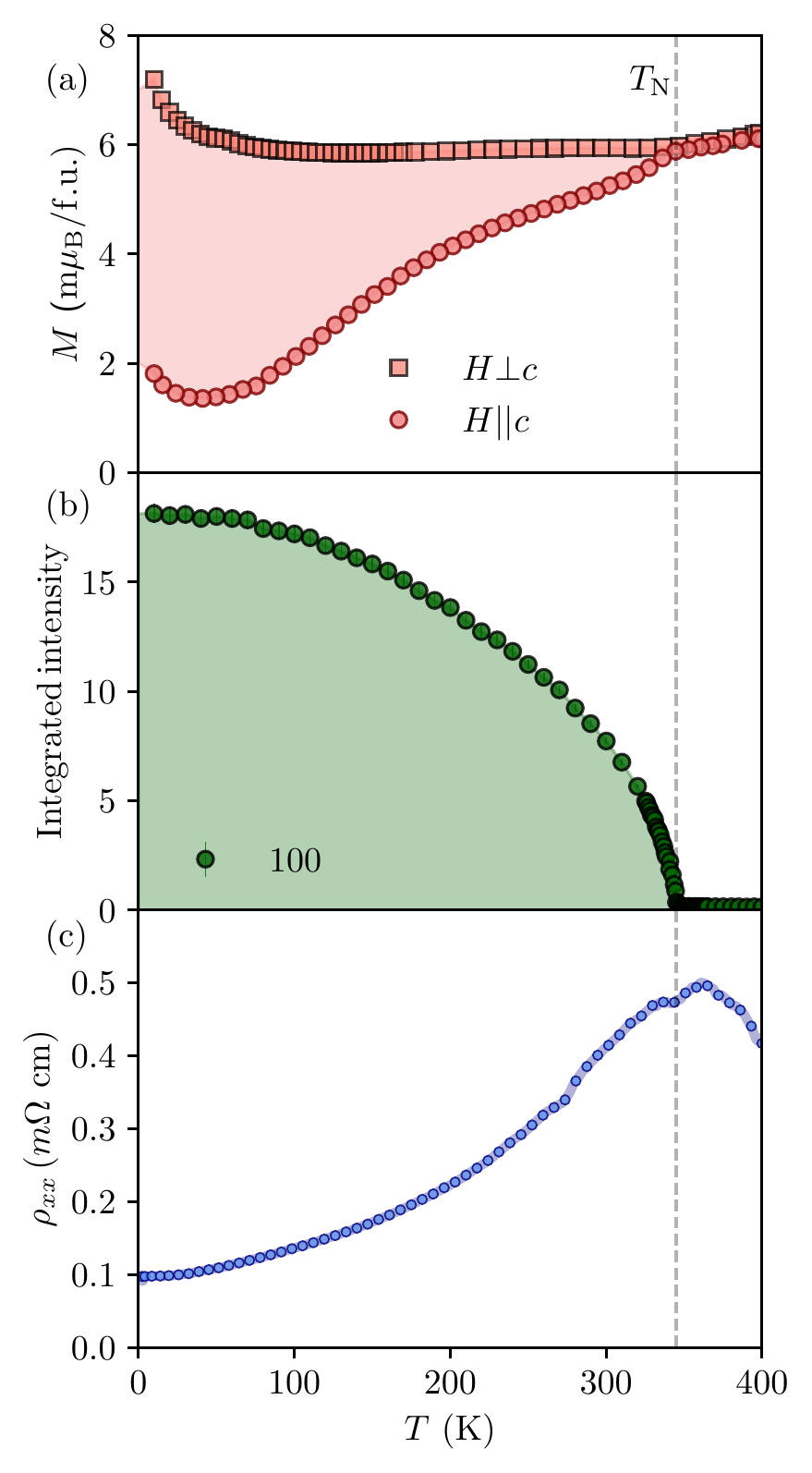}
	\caption{\label{fig:AI0_YbMnSb2_Figure_2.png}(a) Temperature dependent magnetization curves of \yms\, measured with a field $\mu_0H = 3$\,T applied parallel and perpendicular to the crystal $c$ axis. The bifurcation of the curves below $T_\mathrm{N}$\,=\,345(1)\,K indicate the onset of AFM order. (b) Integrated intensity of the 100 magnetic reflection measured by neutron diffraction.  (c) Temperature dependence of the in-plane resistivity of \yms, displaying  metallic conductivity for $T < T_\mathrm{N}$.}
\end{figure}

Polarized neutron diffraction measurements ($\lambda = 0.832$\,{\AA}) were performed on the single-crystal diffractometer D3 (ILL) with a polarized incident beam but without polarization analysis (half-polarized setup). The incident neutron polarization ($P_\mathrm{i}$) was oriented either parallel or antiparallel to $z$ (the normal to the scattering plane) by means of a cryo-flipper, and maintained by guide fields. The crystal was aligned inside a vertical field superconducting magnet with the crystal $b$ axis vertical to give access a series of reflections in the $h0l$ horizontal scattering plane [Fig.~\ref{fig:AI0_YbMnSb2_Figure_6.png}(b)]. A magnetic field of $\mu_0H = 0.4$\,T was applied.

For both neutron diffraction experiments on the instruments D3 and D10, the single crystals were pre-aligned with the neutron Laue diffractometers OrientExpress~\cite{OULADDIAF20061052} and CYCLOPS~\cite{Ouladdiaf2011} at ILL. Neutron diffraction data analysis was performed with the {\it Mag2Pol} software~\cite{QureshiMag2Pol}.

To clarify the topological nature of the electronic band structure in \yms\, we performed density functional theory (DFT) calculations of the electronic band structure using the Quantum Espresso implementation~\cite{Giannozzi_2009}. Relativistic pseudo-potentials were used in the calculations to account for the large spin-orbit coupling (SOC) arising from the heavy Sb ions which might lead to band inversion. Furthermore, a Hubbard $U = 4.1$ eV was used to model the strong electron-electron correlations due to the Mn $3d$ bands. The Mn spin configuration used in the calculations was obtained from the analysis of the neutron scattering data as input parameters. A Monkhorst–Pack \textbf{k}-point sampling mesh of $4\times4\times6$ was used~\cite{PhysRevB.13.5188}.

\section{Results and Discussion}

Figure~\ref{fig:AI0_YbMnSb2_Figure_2.png}(a) plots the magnetization of \yms\, as a function of temperature in a fixed field of $\mu_0H=$ 3\,T, applied parallel and perpendicular to the crystal $a$ axis. On cooling, we observe a bifurcation of the curves below $T_{\rm N}$ = 345(1)\,K, indicating the onset of antiferromagnetic order of the Mn spins (the Yb atoms are divalent and therefore non-magnetic). The anisotropy of the induced moment in the magnetically ordered phase, with $M_{\perp c} > M_{\| c}$, suggest that the crystal $c$ axis is an easy axis. At low temperatures (below $T$ = 50\,K), we observe an upturn in the magnetization in both field configurations, consistent with an earlier report~\cite{Wang2018}. In a later section, we will discuss the possible origin of this behavior.

Figure~\ref{fig:AI0_YbMnSb2_Figure_2.png}(b) displays the temperature dependence of the integrated intensity of the 100 reflection measured by neutron diffraction.  The 100 reflection is structurally forbidden by the $n$ glide, and its intensity has an order parameter-like dependence. Its appearance coincides with the bifurcation in the magnetic susceptibility at $T_\mathrm{N}$. These observations confirm the existence of antiferromagnetism in \yms. 

In Fig.~\ref{fig:AI0_YbMnSb2_Figure_2.png}(c)  we plot the in-plane longitudinal resistivity ($\rho_{xx}$)  as a function of temperature. On cooling, we observe that $\rho_{xx}$ increases to a maximum at $T_\mathrm{N}$ then decreases with temperature down to $T$ = 2\,K (the lowest temperature reached) in good agreement with earlier studies \cite{Kealhofer2018,Wang2018,Qiu2019}.

The broad resistivity peak in the vicinity of $T_\mathrm{N}$ could indicate a coupling between charge transport and  fluctuations of the Mn moments, which diverge at $T_\mathrm{N}$.  \textit{Ab inito} calculations of the electronic band structure of \yms\, demonstrate that the Sb square layer, which is structurally sandwiched between two Mn layers [Fig.~\ref{fig:AI0_YbMnSb2_Figure_1.png}(a)], is mainly responsible for charge transport~\cite{Qiu2019}. Coupling between Mn $3d$ and Sb $4p$ states makes it possible for different Mn magnetic structures to cause different Sb $4p$ electronic topologies [see Figs.~\ref{fig:AI0_YbMnSb2_Figure_1.png}(b)--(d)], and this is why it is important to understand the details of the AFM structure observed below $T_{\rm N}$.

All of the magnetic structures depicted in Fig.~\ref{fig:AI0_YbMnSb2_Figure_1.png} can be described with a magnetic propagation vector $\textbf{q} = (0,0,0)$. For $T < T_{\rm N}$, therefore,  magnetic Bragg peaks observed by neutron diffraction may coincide with structural Bragg peaks. Figure~\ref{fig:AI0_YbMnSb2_Figure_4.pdf}(a) illustrates the set of reflections measured in our diffraction study. These reflections are mostly structurally allowed in the $P4/nmm$ space group (indicated by the blue circles), but there are also a small number of reflections of the form $h + k = 2m+1$ ($m$ integer)  which are forbidden by the $n$ glide (red squares). The latter reflections, if present, are therefore purely magnetic in origin.
%{\color{magenta} Given that all of the predicted magnetic configurations in Figs~\ref{fig:AI0_YbMnSb2_Figure_1.png}(b)--(d) can be described by the magnetic propagation vector $\textbf{q} = (0,0,0)$, we would expect the magnetic reflections obtained from single crystal neutron diffraction to be commensurate with the peaks arising from the host $P4/nmm$ crystal structure.} 

To identify the magnetic contribution to the structural Bragg peaks when $T < T_{\rm N}$ it is important to obtain a good model for the crystal structure in the paramagnetic phase. Figure~\ref{fig:AI0_YbMnSb2_Figure_4.pdf}(b) shows the results of our refinement in the paramagnetic phase at $T$ = 400\,K. The observed and calculated peak intensities are seen to agree well. The refined structural parameters are given in the Supplemental Material~\cite{YbMnSb2Supp}.  %The $a$ axis of the orthorhombic cell is chosen to be perpendicular to the square sublattice of Sb atoms, and since the $b$ and $c$ cell parameters have very similar lengths, twinning due to a $90^\circ$ rotation about $a$ is allowed~\cite{Parsons:ba5036} and was accounted for in the refinement. The population of each twin was determined to be 47.3\% and 52.7\%, which is close to 50\% as expected since neither twin should be energetically favored in the preparation of the crystal. 

For completeness, we mention that we also observed some very weak reflections at half-integer $l$ positions which are structurally forbidden in $P4/nmm$ but allowed in the closely related $Pnma$ space group. The latter space group was reported to give an equally good description of x-ray  diffraction data from polycrystalline \yms~\cite{Wang2018} and is known in the wider family of $A$MnSb$_2$ ($A$ = Eu, Sr, Ca) compounds~\cite{Soh2019b,Gong2020,Liu2017,Liu2017b, He2017,Zhang2020}. These half-integer reflections can arise from a subtle structural distortion which doubles the unit cell in the out-of-plane direction, but  due to limited data we have not attempted to refine the structure further. The central conclusions of the study, which concern the magnetic structure, are not affected by the existence of such a distortion.

 Figures~\ref{fig:AI0_YbMnSb2_Figure_4.pdf}(c)--(e) compare the measured ($I_\mathrm{obs}$) and predicted ($I_\mathrm{calc}$) integrated intensities based on the magnetic structures shown in Figs.~\ref{fig:AI0_YbMnSb2_Figure_1.png}(b), (c) and (e). There are large discrepancies between $I_\mathrm{obs}$ and $I_\mathrm{calc}$ for the G-type AFM model, Fig.~\ref{fig:AI0_YbMnSb2_Figure_4.pdf}(b), especially for the purely magnetic reflections  which have zero intensity for this magnetic structure. Hence, the G-type AFM order, shown in Fig.~\ref{fig:AI0_YbMnSb2_Figure_1.png}(b), can safely be excluded. Similarly, the $I_\mathrm{calc}$ values for the in-plane C-type AFM model significantly underestimate the intensities of the purely magnetic reflections [Fig.~\ref{fig:AI0_YbMnSb2_Figure_4.pdf}(c)]. Therefore, the in-plane C-type AFM model shown in Fig.~\ref{fig:AI0_YbMnSb2_Figure_1.png}(c) can also be excluded. 

\begin{figure}[b!]
	\includegraphics[width=0.4\textwidth]{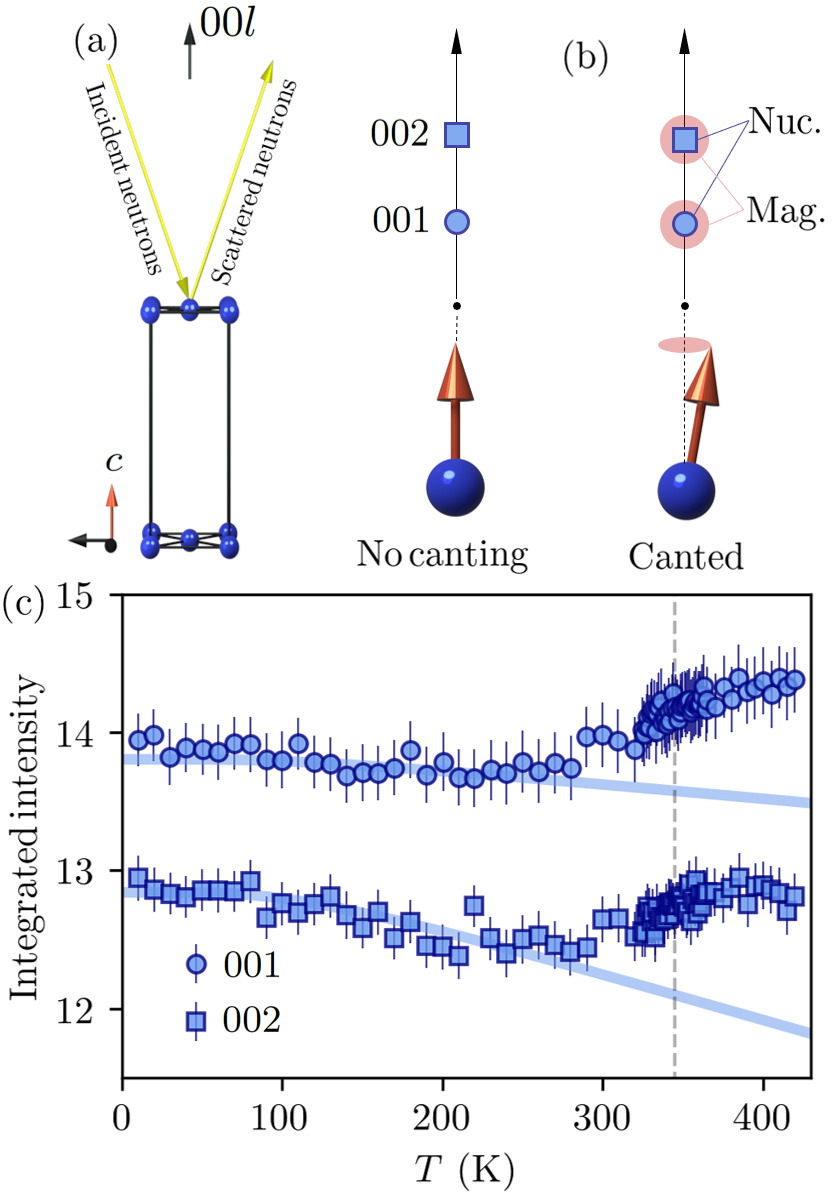}
	\caption{\label{fig:AI0_YbMnSb2_Figure_5.png}(a) Unpolarized neutrons were used on the D10 instrument (ILL), with the scattering vector \textbf{Q} along $h00$. (b) Since magnetic neutron diffraction is sensitive to the component of \textbf{M} perpendicular to \textbf{Q}, namely $\textbf{M}_\perp$, a canting of the Mn moments away from the crystal $c$ axis will produce a magnetic contribution to the nuclear Bragg reflections along $00l$. (c) The temperature dependence of the integrated intensities of the 001 and the 002 reflections. The line corresponds to the fit to the Debye-Waller factor.}
\end{figure}

On the other hand, there is good agreement [see Fig.~\ref{fig:AI0_YbMnSb2_Figure_4.pdf}(d)] between the predicted and measured integrated intensities for the C-type AFM model with Mn moments along the crystal $c$ axis, shown in Fig.~\ref{fig:AI0_YbMnSb2_Figure_1.png}(e). The size of the Mn moments obtained from the refinement is 3.48(9)\,$\mu_\mathrm{B}$, which is somewhat less than the value of 5\,$\mu_\mathrm{B}$ expected for divalent manganese with localized electrons ($3d^5$, $S = 5/2$), but consistent with the Mn ordered moments reported in related $A$Mn$X_2$ intermetallic compounds ($A$ = Ca, Sr, Ba, Yb; $X$ = Sb, Bi)~\cite{Liu2017b,Liu2016,Guo2014,Wang2016,Liu2017c}. 

Let us now consider the structure shown in Fig.~\ref{fig:AI0_YbMnSb2_Figure_1.png}(d), which involves a  canting of the C-type AFM structure so as to produce a net in-plane ferromagnetic (FM) moment. The possibility of a small FM canting was suggested by Wang \textit{et al.}~\cite{Wang2018} to account for the upturn in the magnetization at low temperatures, and was reported in  neutron scattering measurements on the sister compound SrMnSb$_2$~\cite{Liu2017b}. In that study, Liu \textit{et al.} reported an increase in the integrated intensities of the $200$ reflection in the vicinity of $T_\mathrm{N}$, 
and attributed the increase to an in-plane ferromagnetic order which sets in at $T_\mathrm{C} \sim $ 565\,K (well above $T_\mathrm{N}$). This in-plane ferromagnetic order in SrMnSb$_2$ is thought persist into the AFM ordered phase below $T_\mathrm{N}$, giving rise to the canted AFM order shown in Fig.~\ref{fig:AI0_YbMnSb2_Figure_1.png}(d).

To search for a small in-plane FM component we studied the temperature dependence of several $00l$ reflections. Magnetic neutron diffraction is caused by the component of the magnetic moment perpendicular to the scattering vector \textbf{Q}, so if the moments are aligned along the $c$ axis then there no magnetic contribution to any $00l$ reflections. On the other hand, any canting of the moments will produce a magnetic component perpendicular to $c$ which will give magnetic scattering at $00l$ reflections.  Figures~\ref{fig:AI0_YbMnSb2_Figure_5.png}(a) and (b) illustrate the arrangement just described.

Figure~\ref{fig:AI0_YbMnSb2_Figure_5.png}(c) plots the temperature dependence of the integrated intensities of the $001$ and $002$ reflections. At low temperatures (below $T$ = 50\,K), we do not observe an upturn in the integrated intensities which would indicate the onset of in-plane ferromagnetic order. Therefore, the increase in magnetic susceptibility at low temperatures, shown in Fig.~\ref{fig:AI0_YbMnSb2_Figure_2.png}(a), cannot be attributed to the development of spontaneous ferromagnetic component in the crystal $a$-$b$ plane. This conclusion is consistent with behavior of the magnetization as a function of field at low $T$ which also does not indicate spontaneous in-plane ferromagnetism [see Supplemental Materials~\cite{YbMnSb2Supp}]. Instead, we attribute the upturn in the magnetic susceptibility at low temperatures as due to a very small concentration of paramagnetic impurities, as reported in other members of the $A$Mn$X_2$ family~\cite{Soh2019,Guo2014}. Hence, the canted AFM order, shown in Fig.~\ref{fig:AI0_YbMnSb2_Figure_1.png}(d), can be excluded at low temperatures.

\begin{figure}[t!]
	\includegraphics[width=0.4\textwidth]{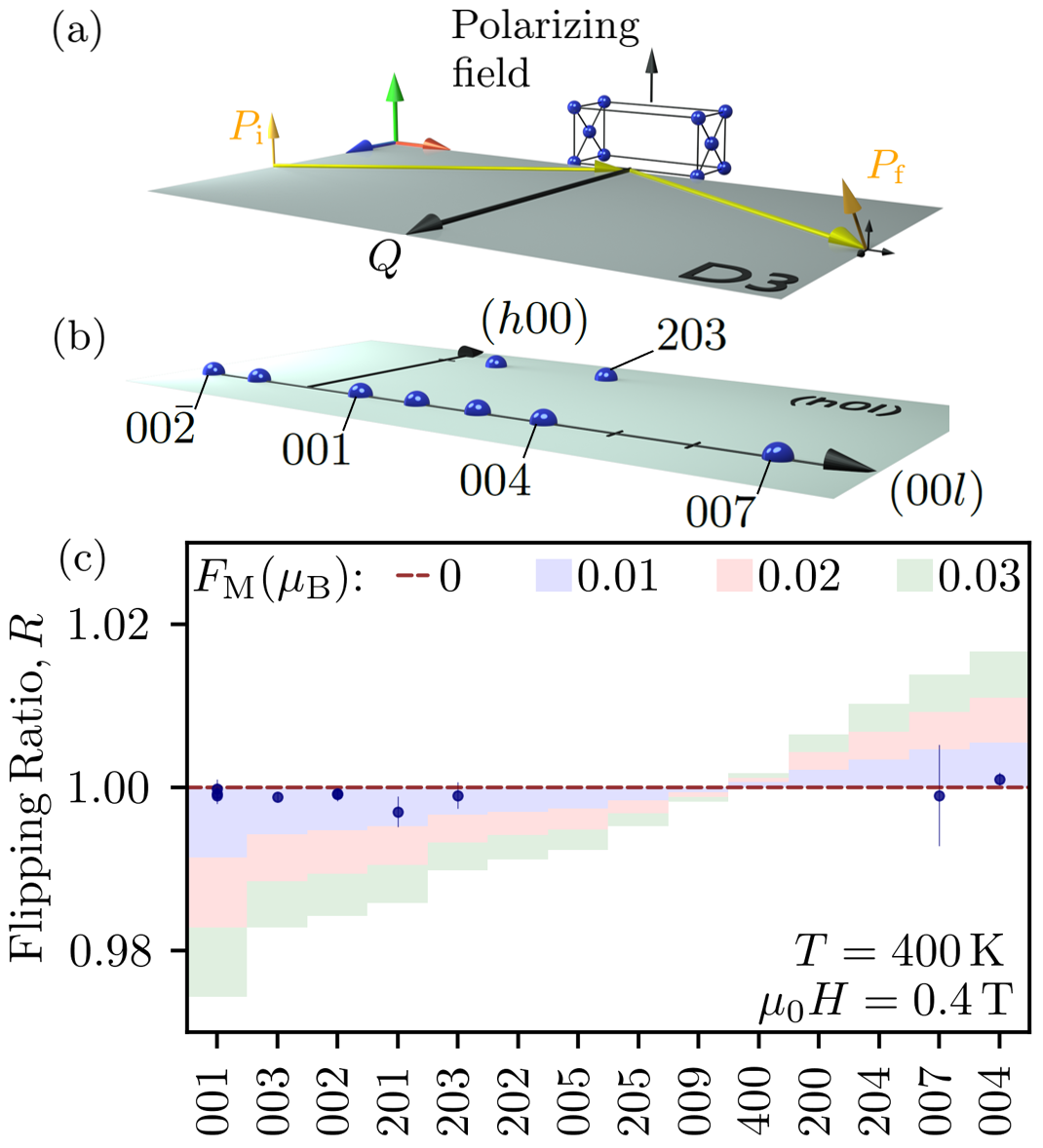}
	\caption{\label{fig:AI0_YbMnSb2_Figure_6.png}(a) To access the $h0l$ reflections in the half-polarized experimental setup on D3, a \yms\, single crystal was aligned with the crystal $b$ axis vertical. The incident neutrons are polarized along $z$  ($P_\mathrm{i} \parallel z$) and the scattered neutron polarization  $P_\mathrm{f}$ is not analyzed. (b) Reciprocal space plot of the measured reflections (blue spheres) in the $h0l$ scattering plane. (c) Measured flipping ratios (blue circles) along with the calculated $R$ for several in-plane magnetic moments from 0\,$\mu_\mathrm{B}$ (dashed line) to 0.03\,$\mu_\mathrm{B}$.}
\end{figure}

However, we observed a gradual increase in the integrated intensity of the 001 and 002 reflections above $T \simeq$  200\,K [Fig.~\ref{fig:AI0_YbMnSb2_Figure_5.png}(c)]. This increase, which is too large to be accounted for by paramagnetic diffuse scattering,  could potentially arise from an in-plane ferromagnetic order of the Mn moments, giving rise to a canted magnetic structure [Fig.~\ref{fig:AI0_YbMnSb2_Figure_1.png}(d)] when $T < T_{\rm N}$, as was reported in the neutron diffraction study of SrMnSb$_2$~\cite{Liu2017b}. To test for the presence of in-plane ferromagnetism in \yms\, we used polarized neutron diffraction to measure the flipping ratio ($R$) of a series of reflections within the $h0l$ scattering plane at $T = 400\,\textrm{K} > T_\mathrm{N}$ [see Fig.~\ref{fig:AI0_YbMnSb2_Figure_6.png}(a) and  (b)].	

For centrosymmetric crystals and a magnetic field that is perpendicular to the scattering plane the flipping ratio is given by~\cite{Boothroyd2020}, 
\begin{equation}
	\label{eqn:flippingratio}
    R = \frac{F_\mathrm{N}^2+F_\mathrm{M}^2+2F_\mathrm{N}F_\mathrm{M}}{F_\mathrm{N}^2+F_\mathrm{M}^2-2F_\mathrm{N}F_\mathrm{M}},
\end{equation}
where $F_\mathrm{M}$ and $F_\mathrm{N}$ are the magnetic and nuclear structure factors. When there is no magnetic contribution to the measured reflections ($F_\mathrm{M} = 0$) the flipping ratio is seen from Eq.~(\ref{eqn:flippingratio}) to be $R = 1$. Conversely, a flipping ratio that deviates from unity implies a non-zero $F_\mathrm{M}$. As the applied field was small ($\mu_0H = 0.4$\,T), the induced moment on paramagnetic Mn is negligible. Hence, if $R \ne 1$ we can infer a small FM contribution.
	
Figure~\ref{fig:AI0_YbMnSb2_Figure_6.png}(c) plots the measured flipping ratios $R$ of the reflections that were studied. The plot also includes the calculated $R$ for several values of the in-plane ferromagnetic moment up to 0.03\,$\mu_\mathrm{B}$. In particular, the dashed horizontal line denotes $R = 1$, which corresponds to zero ferromagnetic moment. 

It can be seen from Fig.~\ref{fig:AI0_YbMnSb2_Figure_6.png}(c) that the measured $R$ is consistent with an in-plane ferromagnetic component of less than 0.01$\,\mu_\mathrm{B}$. This is far too small to explain the observed increase in integrated intensity of the $00l$ reflections above $T \sim$ 200\,K, which we estimate would require an in-plane moment of about 0.8\,$\mu_\mathrm{B}$. Hence, the polarized neutron measurements exclude spontaneous in-plane magnetic order above $T \sim$ 200\,K.  Furthermore, the magnetization as a function of field at $T$ = 400\,K also indicates that any in-plane ferromagnetic component is negligible ($M \lesssim 0.001\,\mu_\mathrm{B}$) (see Supplemental Materials~\cite{YbMnSb2Supp}). Therefore, although it is unclear what causes the observed rise in the integrated intensity of $00l$ reflections above $T \sim $ 200\,K [Fig.~\ref{fig:AI0_YbMnSb2_Figure_5.png}(c)], we can safely exclude a ferromagnetic origin. 

The evidence presented here excludes all of the proposed magnetic structures presented in Figs.~\ref{fig:AI0_YbMnSb2_Figure_1.png}(b)--(d) as acceptable models for the magnetic order of \yms, and instead we find C-type AFM order with Mn moments along the crystal $c$ axis [Fig.~\ref{fig:AI0_YbMnSb2_Figure_1.png}(e)]. We note that some of the predicted magnetic structures in Figs.~\ref{fig:AI0_YbMnSb2_Figure_1.png}(b)--(d) were based on \textit{ab initio} calculations, and the difference between the self-consistent energies obtained for the various magnetic structures can sometimes be of the same order of magnitude as the computational uncertainty. Hence, our work highlights the importance of identifying the ground state magnetic structure experimentally before proceeding to calculate the electronic band structure. 

\begin{figure}[t!]
	\includegraphics[width=0.4\textwidth]{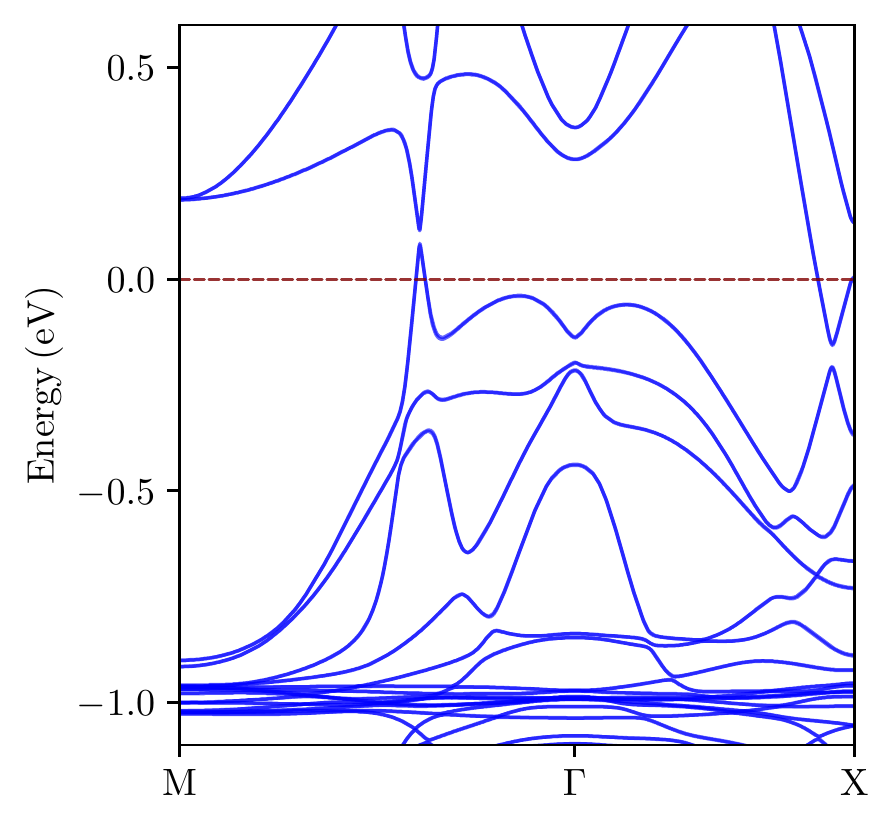}
	\caption{\label{fig:AI0_YbMnSb2_Figure_3.pdf} \textit{Ab initio} electronic dispersion of \yms\, along  in-plane high symmetry directions. The calculation assumes the magnetic structure determined in this study.}
\end{figure}

Finally, Fig.~\ref{fig:AI0_YbMnSb2_Figure_3.pdf} shows the in-plane electronic dispersion of \yms\, calculated with the magnetic structure determined in this study. The  band structure contains two avoided Dirac crossings close the Fermi energy, $\Delta E\sim -0.2$\,eV along $\Gamma$--$X$ high symmetry line and $\Delta E\sim 0.15$\,eV between the $M$ and $\Gamma$ point. These gaps arises from the hybridization of the Sb $5p$ bands~\cite{Klemenz2019,PhysRevLett.107.126402}. The gapped Dirac dispersion of YbMnSb$_2$ is similar to that  reported in other compounds exhibiting C-type AFM magnetic order in the $A$MnSb$_2$ family~\cite{He2017,SciPostPhys.4.2.010,Huang6256,LiuSciRep}.
\section{Conclusion}
In this study we tested  experimentally several magnetic structures which have been predicted for \yms.  Magnetic structure refinements made from unpolarized neutron diffraction data provide conclusive evidence for  C-type AFM order with Mn spins along the $c$ axis.  Polarized neutron diffraction measurements could not detect any in-plane FM order above or below $T_{\rm N}$. The results do not support predictions that \yms\, is either a nodal line or a Weyl semimetal, and instead imply that the band structure contains gapped Dirac nodes near the Fermi level. More generally, this work highlights how different magnetic structures can profoundly affect the topological behavior of electrons in metals.

\begin{acknowledgments}
		The authors  wish  to  thank H. R{\o}nnow for helpful discussions, and D. Prabhakaran for assistance with sample preparation.  The proposal numbers for the neutron scattering experiments on \yms\, are 5-41-1049 (D10), 5-53-305 (D3~\cite{D3_2021}), EASY-671 (CYCLOPS) and EASY-667 (OrientExpress). Y-F.G. and A.T.B. acknowledge support from the Oxford--ShanghaiTech collaboration project. S. M. T. was supported by a scholarship from the Rhodes Trust. J.-R. S. acknowledges support from the Singapore National Science Scholarship, Agency for Science Technology and Research. 
\end{acknowledgments}
\bibliography{ref.bib}
\end{document}